\documentclass[12pt]{article} 

\usepackage{amssymb}
\usepackage{amsbsy}
\usepackage{amsmath}
\usepackage{epsfig}
\usepackage{graphics,color}

\newcommand{\be}{\begin{equation}}
\newcommand{\ee}{\end{equation}}

\newcommand{\bea}{\begin{eqnarray}}
\newcommand{\eea}{\end{eqnarray}}
\newcommand{\bean}{\begin{eqnarray*}}
\newcommand{\eean}{\end{eqnarray*}}

\newcommand{\xv}{{\mathbf x}}

\newcommand{\nn}{\nonumber}

\newcommand{\hm}{\hspace*{-0.6cm}}

\newcommand{\bB}{{\bf B}}
\newcommand{\bD}{{\bf D}}
\newcommand{\bDl}{{\bf\Delta}}
\newcommand{\bE}{{\bf E}}
\newcommand{\bsigma}{{\boldsymbol\sigma}}

\begin{document}

\title{\bf \large Non-relativistic spectrum of two-color QCD at non-zero
  baryon density}

\author{Simon Hands${}^a$, Seyong Kim${}^{a,b}$, Jon-Ivar Skullerud${}^c$ \\
\mbox{} \\
\mbox{} \\
{${}^a$\em\normalsize Department of Physics, College of Science, Swansea University,
  Swansea SA2 8PP, United Kingdom}\\
{${}^b$\em\normalsize Department of Physics, Sejong University, 
Gunja-Dong, Gwangjin-Gu, Seoul 143-747, Korea} \\
{${}^c$\em\normalsize Department of Mathematical Physics, National
  University of Ireland Maynooth,} \\
{\em\normalsize Maynooth, County Kildare, Ireland}
}

\date{\today}

\maketitle

\begin{abstract}

The heavy quarkonium spectrum of Two Color QCD (QC$_2$D) at non-zero quark
chemical potential $\mu$ and temperature $T$ with $\mu/T\gg1$ has been
calculated in both $S$- and $P$-wave channels using a
lattice non-relativistic formulation of QC$_2$D. As $\mu$ is varied,
the quarkonium spectra reveal three separate regions, corroborating
previous findings that there are three distinct physical regimes of
QC$_2$D at low temperature and high baryon density: hadronic matter,
quark/quarkyonic matter, and deconfined matter. The results are interpreted in
terms of the formation of heavy-light $Qq$ states in the two-color baryonic
medium.

\end{abstract}


\newpage

\section{Introduction}
\label{sec:intro}

Cold dense baryonic matter in Quantum ChromoDynamics (QCD) is difficult
to study theoretically because the lattice gauge theory method allowing
a first principles investigation of non-perturbative physics does
not work due to the ``complex action problem'': the introduction of a quark
chemical potential $\mu\not=0$ in the Euclidean formulation of QCD makes its
action complex, and importance sampling used for Monte Carlo evaluation
of the partition function is difficult if not impossible to implement.

In previous work~\cite{Hands:2006ve,Hands:2010gd,Hands:2011ye} we presented
non-perturbative results at non-zero density for QC$_2$D, a QCD-like theory with two
colors, using orthodox lattice simulations. Since the gauge group is SU(2) and
the quark representation pseudoreal, the functional measure 
$\mbox{det}^{N_f}{\cal M}(\mu)$ remains real even once
$\mu\not=0$, and positivity can be ensured by insisting that the number of
flavors $N_f$ is even. In QC$_2$D baryons are bosonic, and hadron
multiplets, including Goldstone boson states associated with the breaking of
global symmetries, contain both $q\bar q$ mesons and $qq$ baryons.
Despite these clear differences from physical QCD, QC$_2$D has an
unexpectedly rich structure as $\mu$ is increased, and we expect many
lessons learned here about issues such as deconfinement, chiral symmetry
restoration, and exotic ground states, may be more widely
applicable.

In brief, we have found three regimes with distinct behaviours beyond the onset
at $\mu=\mu_o\equiv m_\pi/2$ where baryonic matter is first induced into the ground state at
zero temperature. For $\mu_o\leq\mu\leq\mu_Q$ the matter consists of tightly
bound $qq$ scalars, which form a Bose-Einstein Condensate (BEC). Since the
scalar diquarks are Goldstone modes associated with the spontaneous breaking of
global baryon number conservation leading to superfluidity, thermodynamics in
this regime is well-described by an effective approach based on chiral
perturbation theory~\cite{Kogut:2000ek}. However, at the larger densities 
found in the range $\mu_Q\leq\mu\leq\mu_D$, we find the thermodynamic properties 
of the system to be more like those of a degenerate system of quarks having a
well-defined Fermi surface; superfluidity persists, but is now understood as
resulting from a Bardeen-Cooper-Schrieffer (BCS) mechanism involving the condensation
of weakly-bound diquark pairs at opposite points on the surface. Finally, for
the largest densities $\mu\geq\mu_D$ we find color deconfinement via the classic
signal of the non-vanishing expectation of the Polyakov loop. With the
simulation parameters outlined in Sec.~\ref{sec:formulation} below we
estimate $\mu_o\simeq360$MeV, $\mu_Q\simeq530$MeV, and
$\mu_D\simeq850$MeV~\cite{Hands:2010gd}. The regime $\mu_Q\leq\mu\leq\mu_D$ is
particularly interesting, since it resembles ``quark matter'' while remaining
color-confined; as such it is reminiscent of the ``quarkyonic'' phase first
discussed in~\cite{McLerran:2007qj}.

In this work, we study QC$_2$D with both $\mu\not=0$ and temperature $T>0$ 
using correlation functions associated with heavy quarkonium $Q\bar Q$ states,
which offer a sophisticated probe of medium effects.
The behaviour of quarkonia for $T>0$ in QCD is suggested as one of the
signatures for quark-gluon plasma formation
\cite{Matsui:1986dk}. Since the energy required to create a
$Q-\bar{Q}$ pair is much larger than both $\Lambda_{QCD}$ and the
average thermal energy, the 
$Q$ -- $\bar Q$  production rate is dominated by short distance
physics which is $T$-independent. Heavy quarkonium is a low energy $Q\bar Q$ bound state, 
and the thermal medium found in, say, heavy-ion collisions can influence
quarkonium formation. In \cite{Matsui:1986dk}, the authors modelled
the thermal medium using a screening potential $V_{Q\bar Q}$ and showed the
``melting'' of
heavy quarkonium (i.e., non-existence of $Q\bar Q$ bound states above
a certain temperature) by solving the resulting Schr\"odinger equation.
Evidence for suppression of excited bottomomium states at $T>T_c$ in lattice
simulations is given in \cite{Aarts:2010ek,Aarts:2011sm}.

It is important to stress that in this context the heavy quarks are regarded as
test particles and are not in thermal equilibrium. In the current study this
implies that $Q$, $\bar Q$ do {\it not\/} couple directly to the chemical potential $\mu$.
To treat the heavy quarks we will use an effective approach known as
non-relativistic QCD (NRQCD); in QCD this is usually applied to the
$b$ - $\bar b$ system (the $\Upsilon$ and $\chi_b$ meson families)
where there is a well-defined hierarchy of 
scales. For heavy quark mass $M$ and velocity $v$, for NRQCD to be applicable we
require~\cite{Brambilla:2010cs}
\begin{equation}
M\gg p\sim r^{-1}\sim Mv \gg \Delta E\sim Mv^2,
\label{eq:hierarchy}
\end{equation}
where $\Delta E$ is the quarkonium ``binding energy'' and $r$ is the
typical distance between the quark and the antiquark. For $\Upsilon$ systems
$v^2\sim0.1$. In Sec.~\ref{sec:formulation} we briefly 
review the lattice approach to QC$_2$D with $\mu\not=0$ and outline the formulation of
NRQC$_2$D. Our main results, for spin-singlet and spin-triplet states in both
$S$- and $P$-waves are presented in Sec.~\ref{sec:results}, and a discussion of
the observed $\mu$- and $T$-variation follows in Sec.~\ref{sec:rampant_speculation}. 

%

\section{Formulation}
\label{sec:formulation}

We investigate the heavy quarkonium spectrum at non-zero temperature and 
baryon density 
by calculating ${\cal
O} (v^4)$ non-relativistic QC$_2$D correlators using background
lattice gauge field configurations generated on $16^3\times12,
12^3\times16$, and $12^3\times24$ lattices, at
$\beta = 1.9, \kappa = 0.168$ with two dynamical flavors of Wilson 
quark \cite{Hands:2006ve,Hands:2010gd,Hands:2011ye}). These
parameters correspond to lattice spacing $a = 0.186(8)$ fm ($ =
1/1.060(45)$ GeV$^{-1}$), $m_\pi a = 0.68(1)$ and $m_\pi/m_\rho =
0.80(1)$, where the scale is set by the string tension ((440 MeV)$^2$
at $\mu = 0$). The corresponding temperatures are $T=44,66$ and 88MeV. 
The range of chemical potential studied is $0\le\mu a\le1.1$; in
\cite{Hands:2010gd} at $\mu=0.8a^{-1}\approx850$MeV a quark density $n_q=16$ -
32fm$^{-3}$ was reported, corresponding to between 35 and 70 times matter
density, where the uncertainty is due to discretisation artifacts.

A standard Hybrid Monte-Carlo algorithm
was used to generate lattice
configurations, where the action
\be
S = \sum_{x,i} \overline{\psi}_i (x) {\cal M}_{x,y}(\mu) \psi_i(y) + \kappa j \sum_x
\left[\psi_{2}^{\rm tr}(x) (C \gamma_5) \tau_2 \psi_1(x) - h.c.\right] 
\ee
with
\be
{\cal M}_{x,y} = \delta_{x,y} - \kappa \sum_\nu \left[(1-\gamma_\nu)e^{\mu
    \delta_{\nu,0}} U_\nu (x) \delta_{y,x+\nu} + (1+\gamma_\nu)e^{-\mu
    \delta_{\nu,0}} U_\nu^\dagger(y) \delta_{y,x-\nu}\right] .
\ee
The diquark source term proportional to $j$ mitigates large infrared
fluctuations in a superfluid phase with $\langle\psi^{\rm
tr}_2(C\gamma_5)\tau_2\psi_1\rangle\not=0$, and also helps ergodicity by
enabling real eigenvalues of ${\cal M}$ to traverse the origin.
To assess the effect of the diquark source, 
configurations generated with two different magnitudes 
$j = 0.02$ and $0.04$ were used. Details of
the simulation algorithm and previous analyses of various
bulk thermodynamic quantities are
given in \cite{Hands:2006ve,Hands:2010gd,Hands:2011ye} and the
temperature dependence of $\mu_D$ is discussed in \cite{Giudice:2011zu,Cotter:2012}. 

We used the following ${\cal O}(v^4)$ Euclidean NRQC$_2$D lagrangian
density for the heavy quark with mass $M$:
\be
\label{eq:LNRQCD}
{\cal L}_Q = {\cal L}_0 + \delta {\cal L},
\ee
with
\be
\label{LNRQCD_1}
{\cal L}_0 = \phi^\dagger \left(D_\tau - \frac{\bD^2}{2M} \right) \phi +
\chi^\dagger \left(D_\tau + \frac{\bD^2}{2M} \right) \chi,
\ee
and
\bea
\label{eq:LNRQCD_2}
\delta {\cal L} = &&\hm - \frac{c_1}{8M^3} \left[\phi^\dagger
  (\bD^2)^2 \phi - \chi^\dagger (\bD^2)^2 \chi \right] \nonumber
\\ &&\hm + c_2 \frac{ig}{8M^2}\left[\phi^\dagger \left(\bD\cdot\bE -
  \bE\cdot\bD\right) \phi + \chi^\dagger \left(\bD\cdot\bE -
  \bE\cdot\bD \right) \chi \right] \nonumber \\ &&\hm - c_3
\frac{g}{8M^2}\left[\phi^\dagger
  \bsigma\cdot\left(\bD\times\bE-\bE\times\bD\right)\phi +
  \chi^\dagger \bsigma\cdot\left(\bD\times\bE-\bE\times\bD\right)\chi
  \right] \nonumber \\ &&\hm - c_4 \frac{g}{2M} \left[\phi^\dagger
  \bsigma\cdot\bB \phi - \chi^\dagger \bsigma\cdot\bB \chi \right]
\eea 
It is similar to 
NRQCD~\cite{Bodwin:1994jh} with the only difference that
$D_\tau$ and $\bD$ are now gauge covariant temporal and spatial
derivatives for SU(2) gauge theory. Here $\phi$ and $\chi$ are
two-color two-spinor fields for the heavy quark
and anti-quark, and the tree-level value for the $c_i$ is 1.

We use the following discretised Green function of (\ref{eq:LNRQCD})
to calculate the heavy quark Green function:
\bea 
G (\xv, \tau=0) = &&\hm S(\xv), \nn\\ G (\xv, \tau=a_\tau) =
&&\hm \left(1 - \frac{H_0}{2n}\right)^n U_4^\dagger(\xv, 0) \left(1 -
\frac{H_0}{2n}\right)^n G(\xv,0), \nn \\ G (\xv, \tau+a_\tau) = &&\hm
\left(1 - \frac{H_0}{2n}\right)^n U_4^\dagger(\xv, \tau) \left(1 -
\frac{H_0}{2n}\right)^n \left(1 -\delta H_n \right) G (\xv, \tau), \eea
where $S(\xv)$ is the source and the lowest-order hamiltonian reads (see
(\ref{eq:derivs}) below)
\be
H_0 = - \frac{\Delta^{(2)}}{2M}, \ee and \bea \delta H_n = &&\hm -
\frac{(\Delta^{(2)})^2}{8 M^3} + \frac{ig}{8 M^2} (\bDl\cdot \bE -
\bE\cdot \bDl) - \frac{g}{8 M^2} \bsigma \cdot (\bDl\times \bE -
\bE\times \bDl) \nonumber \\ &&\hm - \frac{g}{2 M} \bsigma\cdot\bB +
\frac{a^2\Delta^{(4)}}{24 M} - \frac{a(\Delta^{(2)})^2}{16 n M^2}.
\label{eq:deltaH}
\eea 

This approach is used for heavy quarkonium spectroscopy in QCD at $T=0$
in \cite{Davies:1994mp,Davies:1995db,Davies:1998im} and
recently in QCD at $T>0$
\cite{Aarts:2010ek}. The integer $n$
controls the high-momentum behaviour of the evolution equation. Since
in QC$_2$D we do not have any phenomenologically compelling choice for the heavy
quark mass, we choose $Ma \ge 3$ and $n =
1$, to ensure compatibility with (\ref{eq:hierarchy}). 
The last two terms in $\delta H$ are corrections to the kinetic
energy term at non-zero lattice spacing \cite{Lepage:1992tx}. The
lattice covariant derivatives are defined by
\bea 
\Delta_i\phi &=& \frac{1}{2a}\left[ U_i(x)\phi(x+\hat\imath) -
  U_i^\dagger(x-\hat\imath)\phi(x-\hat\imath) \right], \nn
\\ \Delta^{(2)}\phi &=& \sum_i \Delta_i^{(2)} \phi = \sum_i
\frac{1}{a^2}\left[U_i (x) \phi (x+\hat{\imath}) - 2 \phi (x) +
  U_i^\dagger (x - \hat{\imath}) \phi (x - \hat{\imath})\right], \nn
\\ \Delta^{(4)}\phi &=& \sum_i (\Delta_i^{(2)})^2 \phi, \label{eq:derivs}
\eea 
and $\bE$ and $\bB$ in Eq.(\ref{eq:deltaH}) are lattice cloverleaf definitions of
the SU(2) chromoelectric and chromomagnetic fields. To mitigate quantum
corrections we use 
tadpole improvement \cite{Lepage:1992xa}, replacing $ U_\mu (x) \rightarrow
u_0^{-1} U_\mu (x)$, where $u_0$ is the average link determined
from the plaquette expectation value, and setting the coefficients
$c_i$ of (\ref{eq:LNRQCD_2}) to $1$. Note that in SU(2) gauge theory, heavy quarkonium
($Q\bar Q$) states are equivalent to heavy baryon states
($QQ$)~\cite{Hands:1999md}. Since the $P$-wave excitation needs an extended source, 
Coulomb gauge fixing is performed on SU(2) gauge fields
prior to calculation of $G(\xv,\tau)$.
By combining non-relativistic heavy
quark correlators, $^1S_0$, $^3S_1$, $^1P_0$ and $^3P_1$ heavy quarkonium
states could be studied. The expressions for the interpolating operators for all the states are listed in \cite{Thacker:1990bm}.


\section{Results}
\label{sec:results}

\subsection{$S$-wave states}
\begin{figure}
\begin{center}
\includegraphics*[width=8.0cm]{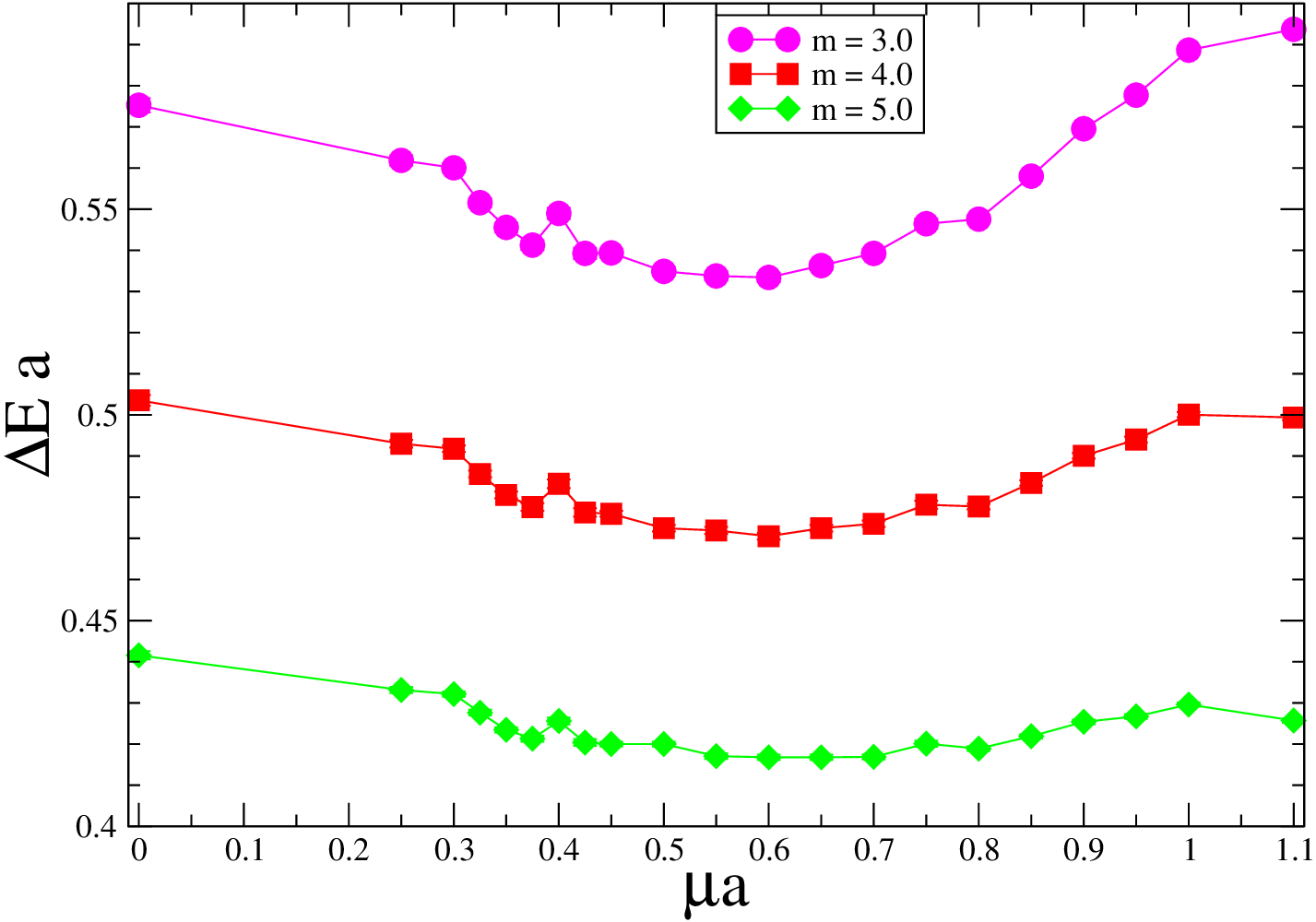}
\includegraphics*[width=8.0cm]{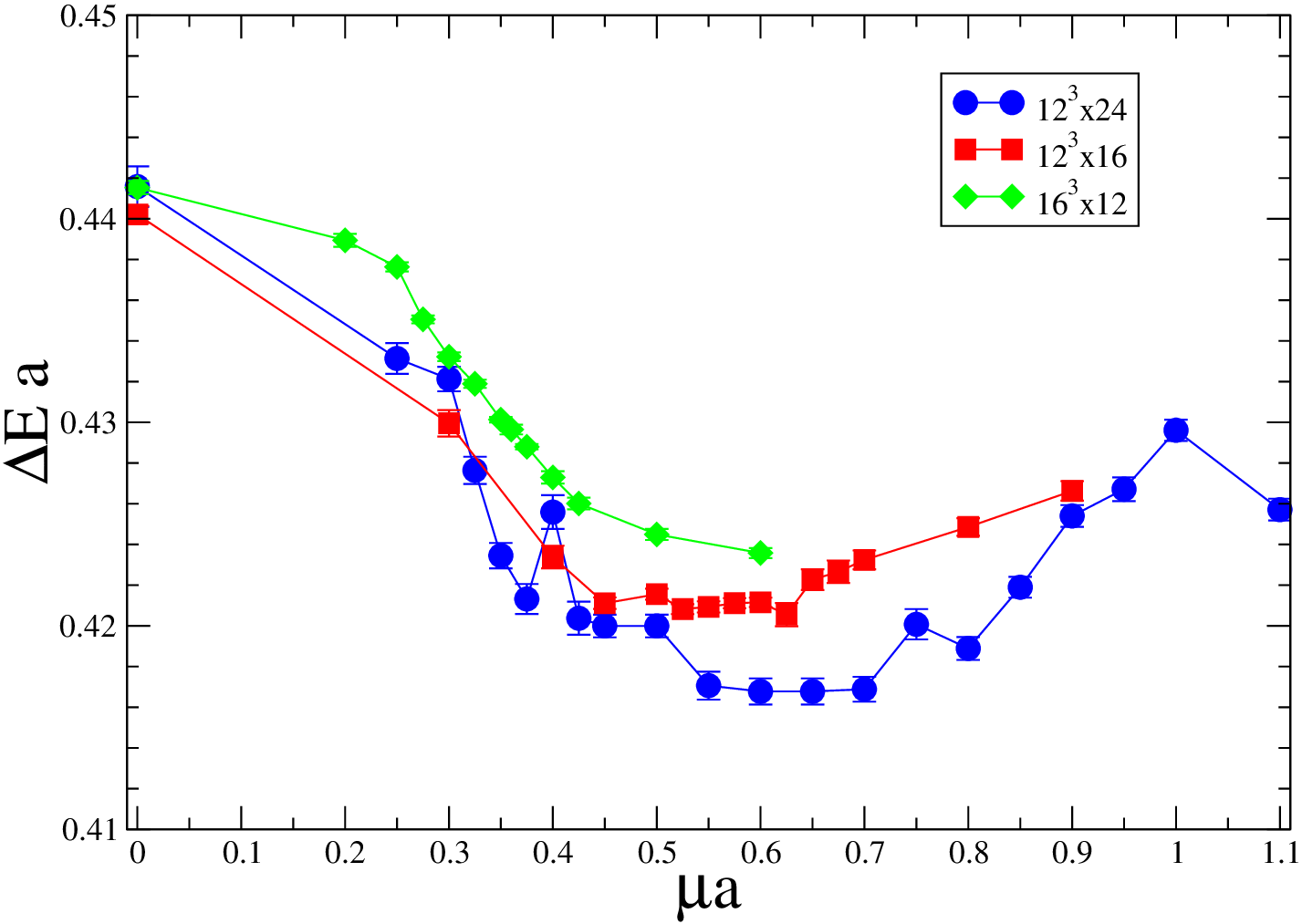}
\end{center}
\caption{(left) Energy of the $^1S_0$ state vs. quark chemical potential $\mu$
  for heavy quark mass $Ma= 3.0, 4.0$ and $5.0$ with $j = 0.02$ on
  $12^3 \times 24$ lattice; (right) Temperature dependence of the
  $^1S_0$ state energy vs. $\mu$ for $Ma = 5.0$ with $j = 0.04$.}
\label{fig:1S0muT} 
\end{figure}
We found that correlators for the S-wave states
$^1S_0$ and $^3S_1$ could be fitted with an exponential decay $\propto
e^{-\Delta E_n\tau}$ corresponding to
a simple pole even once $\mu\not=0$; moreover the fits 
were quite stable over large ranges of $\tau$. This
suggests that $S$-wave quarkonium bound states persist throughout the region
$0 \le \mu a \le 1.1$ and $\frac{1}{24}
\le T a \le \frac{1}{12}$. 

In non-relativistic QCD, $M_n = 2 (Z_M
M - E_0) + \Delta E_n $ for the state $n$, where $Z_M$ is the heavy quark mass
renormalization, $E_0$ is a state-independent additive renormalisation, and $\Delta E_n$ is
the fitted energy of the state \cite{Lepage:1992tx}. Usually, the
experimental value for one of the heavy quarkonium masses is
chosen to fix $E_0$ which is independent of $n$.
This cannot be done in QC$_2$D due to the lack of experimental
spectrum data. However, since introducing $\mu\not=0$, $T>0$
does not induce any new UV
divergences, the change of the $S$-wave state energy from that at $\mu =
0$ or $T = 0$, which must reflect underlying physics, can
be measured.

Fig.~\ref{fig:1S0muT} shows the $T$- and $\mu$-dependences 
of the $^1S_0$
state energy $\Delta E$. The absolute values
have unquantified contributions $E_0(M)$, which in principle could be
subtracted by matching $\Delta E(\mu=0)$, and as such contain little
useful information. As argued above, however, the variation with $\mu$ is
physical.
As $Ma$ increases, and hence higher order effects in $v^2$ become less important, Fig.~\ref{fig:1S0muT} suggests three distinct regimes as $\mu$ is
varied: initially the $^1S_0$ state energy decreases from that at $\mu = 0$,
but once $\mu$ reaches the region $\mu_1 (\simeq 0.5 )\le \mu
a \le \mu_2 (\simeq 0.85)$, the $^1S_0$ state energy stays roughly
constant. For $\mu > \mu_2$, the $^1S_0$ state energy starts
increasing again. 
The variation with $\mu$ becomes more marked as the heavy quark mass $M$ is
decreased; this may possibly be associated with the increasing size of the quarkonium state.
The existence of three distinct
$\mu$-regions in which the $S$-wave quarkonium state
energies show markedly different behaviour and the agreement of
$\mu_{1,2}$ with the values $\mu_Q$ and $\mu_D$ found in \cite{Hands:2010gd}
strengthens the argument for the existence of three different regimes
described in Sec.~\ref{sec:intro} as the BEC
phase, the BCS/quarkyonic phase and the
deconfined phase. 

The three temperatures shown in Fig.~\ref{fig:1S0muT}
all lie below the estimated deconfining transition
temperature $T_c a \sim\frac{1}{6}$ at $\mu = 0$.  As $\mu$ is
increased, there is a deconfinement transition, signalled by the
Polyakov loop increasing from zero, at $\mu_D(T)a\approx0.3,0.55,0.75$
for $Ta=\frac{1}{12},\frac{1}{16},\frac{1}{24}$ respectively
\cite{Giudice:2011zu,Cotter:2012}.  In accordance with this, we see that the $^1S_0$
energy starts increasing roughly at the deconfinement transition at
$Ta=\frac{1}{24}, \frac{1}{16}$.  However, no such behaviour is
observed for the highest temperature.  Interestingly, at $\mu a =
0.6$, 
$ \Delta E a = 0.4157(6), 0.4226(2),
0.4235(2)$  as $Ta$ rises from  $\frac{1}{24}$  to $\frac{1}{12}$. This
positive shift is similar to that observed in the thermal mass of heavy
quarkonium with increasing temperature in hot
QCD with $\mu=0$~\cite{Brambilla:2010vq,Aarts:2011sm}.

\begin{figure}
\begin{center}
\includegraphics*[width=13cm]{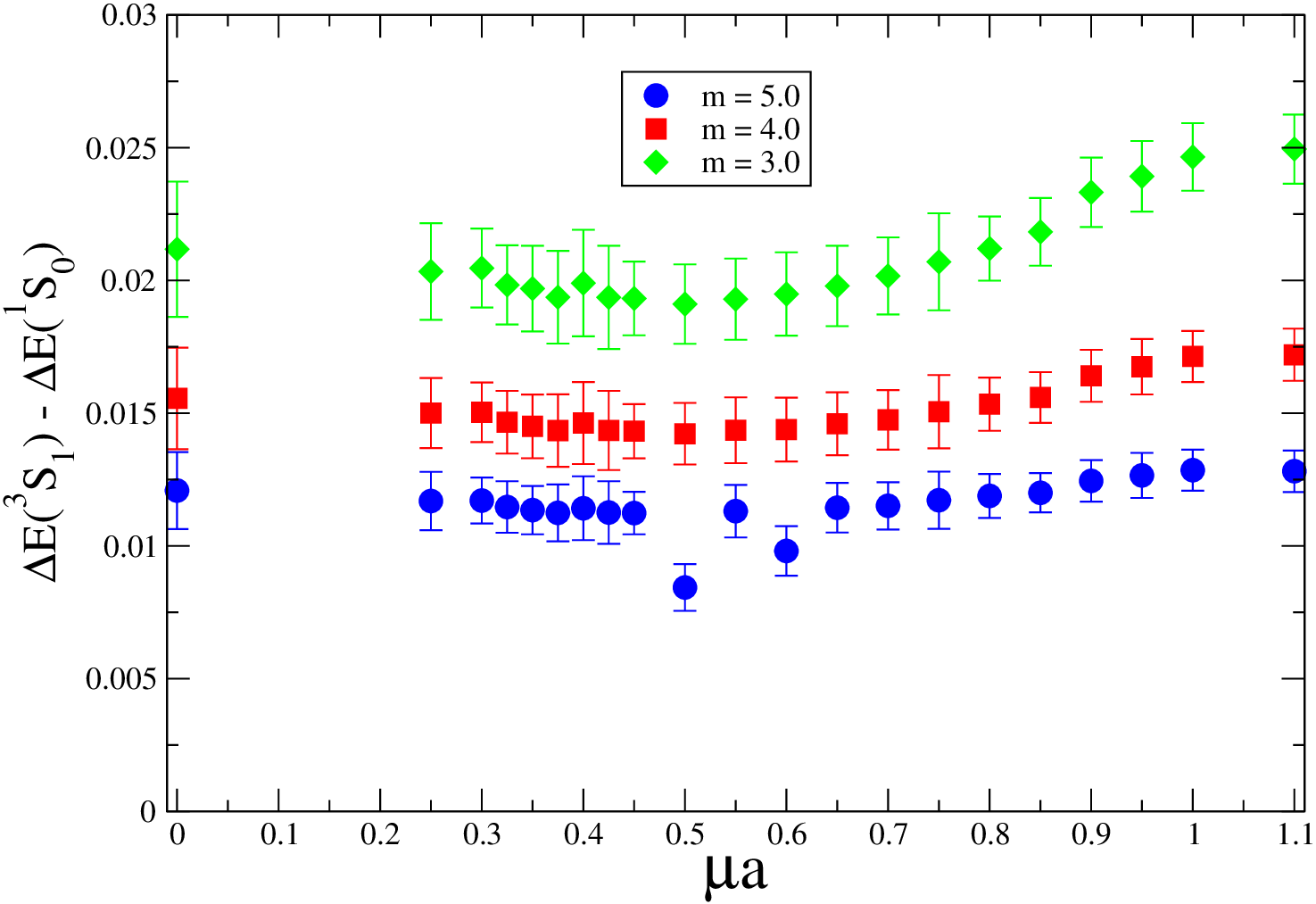}
\end{center}
\caption{The splitting between the $^3S_1$ state energy and $^1S_0$
  state energy for three different $M$ on $12^3 \times 24$}
\label{fig:hf}
\end{figure}
The overall qualitative behavior of the $^3S_1$ state energy is quite
similar to that of the $^1S_0$ state energy. The $^3S_1$ state energy
also shows the same three chemical potential regimes as the $^1S_0$
state energy: the $^3S_1$ state energy decreases until $\mu$
reaches $\mu_1$ and then stays roughly constant until $\mu$ reaches
$\mu_2$ and then increases for $\mu\ge \mu_2$. Thus, instead
of the absolute $^3S_1$ state energy, the hyperfine splitting
$\Delta E_{^3S_1}-\Delta E_{^1S_0}$ is shown as a function of $\mu$
in Fig.~\ref{fig:hf}. Only a weak 
$\mu$-dependence in the splitting for three explored $M$s is observed, at most roughly 
10\% of the magnitude of the effect seen in Fig.~\ref{fig:1S0muT}. This
may not be surprising since even in NRQCD at $T = \mu=0$, the
hyperfine splitting is strongly affected by light quark dynamics
and renormalization effects~\cite{Hammant:2011bt}, and the light
quark mass in our simulation is relatively heavy ($m_\pi a =
0.68(1)$). Clearly, much further study is needed to isolate chemical
potential/temperature effects in the $^3S_1$ - $^1S_0$ splitting.

\begin{figure}
\begin{center}
  \includegraphics*[width=13cm]{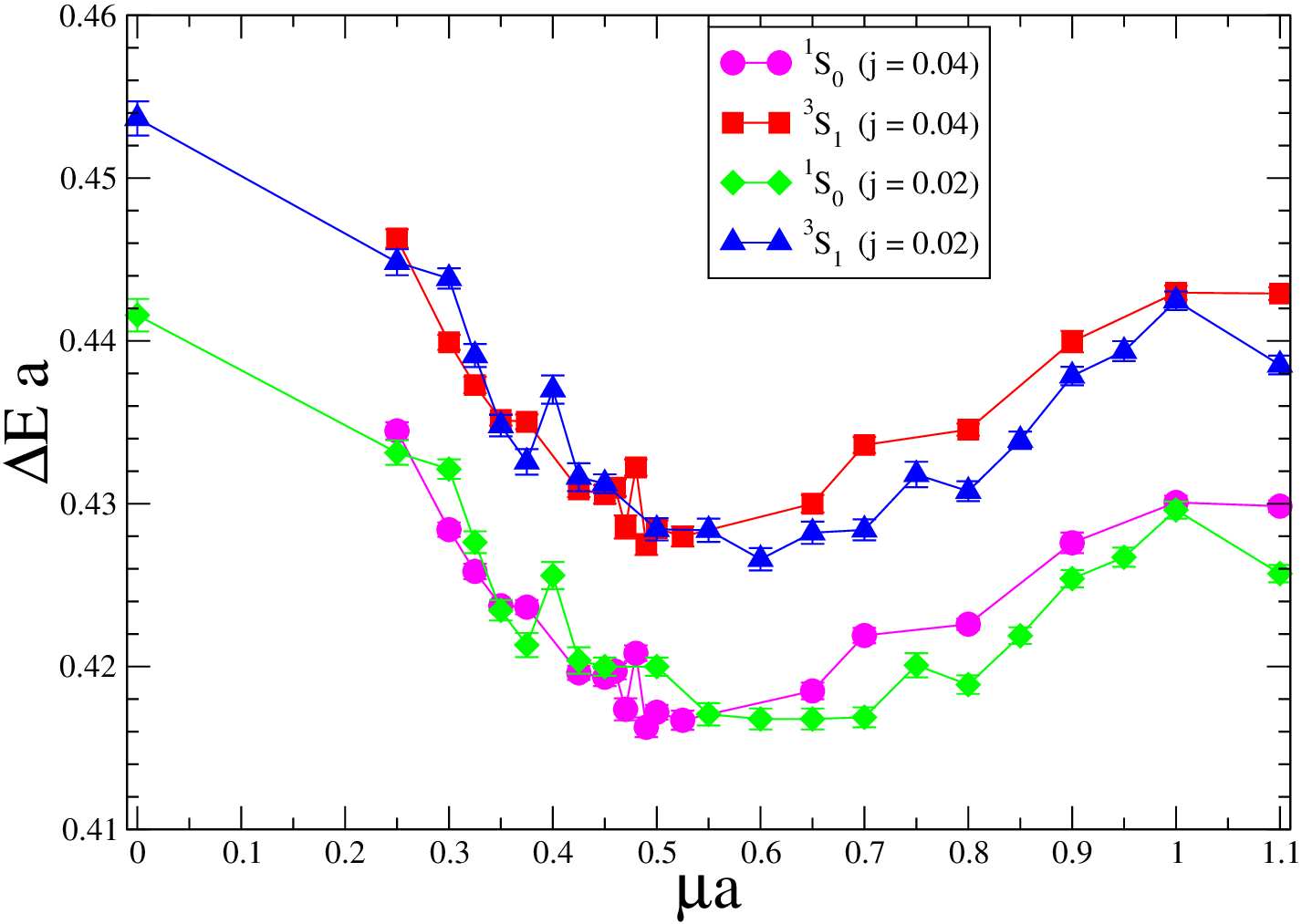}
\end{center}
  \caption{diquark source dependence of the $^1S_0$ and the $^3S_1$
    state energies vs. $\mu$ with $Ma = 5.0$}
\label{fig:qq}
\end{figure}
The influence of the diquark source term 
on the heavy quarkonium spectrum and its effect
in the three distinct $\mu$-regimes we identified
need to be studied. In general, increasing $j$ 
induces a larger superfluid condensation \cite{Hands:2006ve}. How this
affects the heavy quarkonium spectrum is a highly non-trivial
question. Thus, the spectrum calculation has been repeated with
two different source magnitudes $j = 0.02$ and $0.04$.
Fig.~\ref{fig:qq} shows the $j$-dependence of $\Delta E_{^1S_0}$
and $\Delta E_{^3S_1}$. For both $j = 0.02$ and
$0.04$, the $S$-wave state energies continue to manifest three separate regimes, 
but while for $\mu < \mu_1$, the 
$S$-wave state energies show little $j$-dependence, for 
$\mu \ge \mu_1$, the $S$-wave state energies for $j = 0.04$
are mostly larger than those for $j = 0.02$, signifying a larger
diquark condensate effect in the quarkyonic and deconfined regions.

\subsection{$P$-wave states}

\begin{figure}
\begin{center}
\includegraphics*[width=8.0cm]{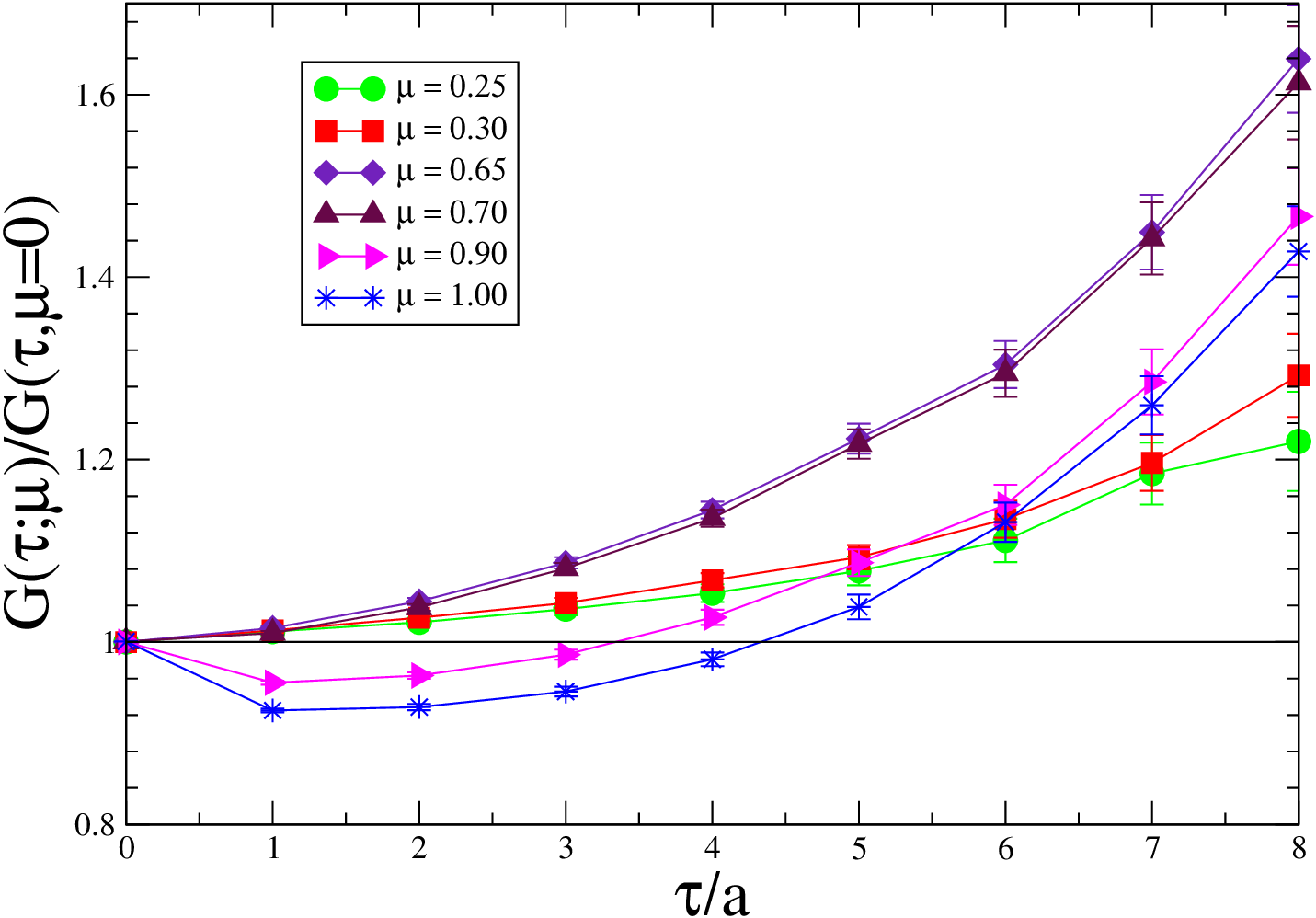}
\includegraphics*[width=8.0cm]{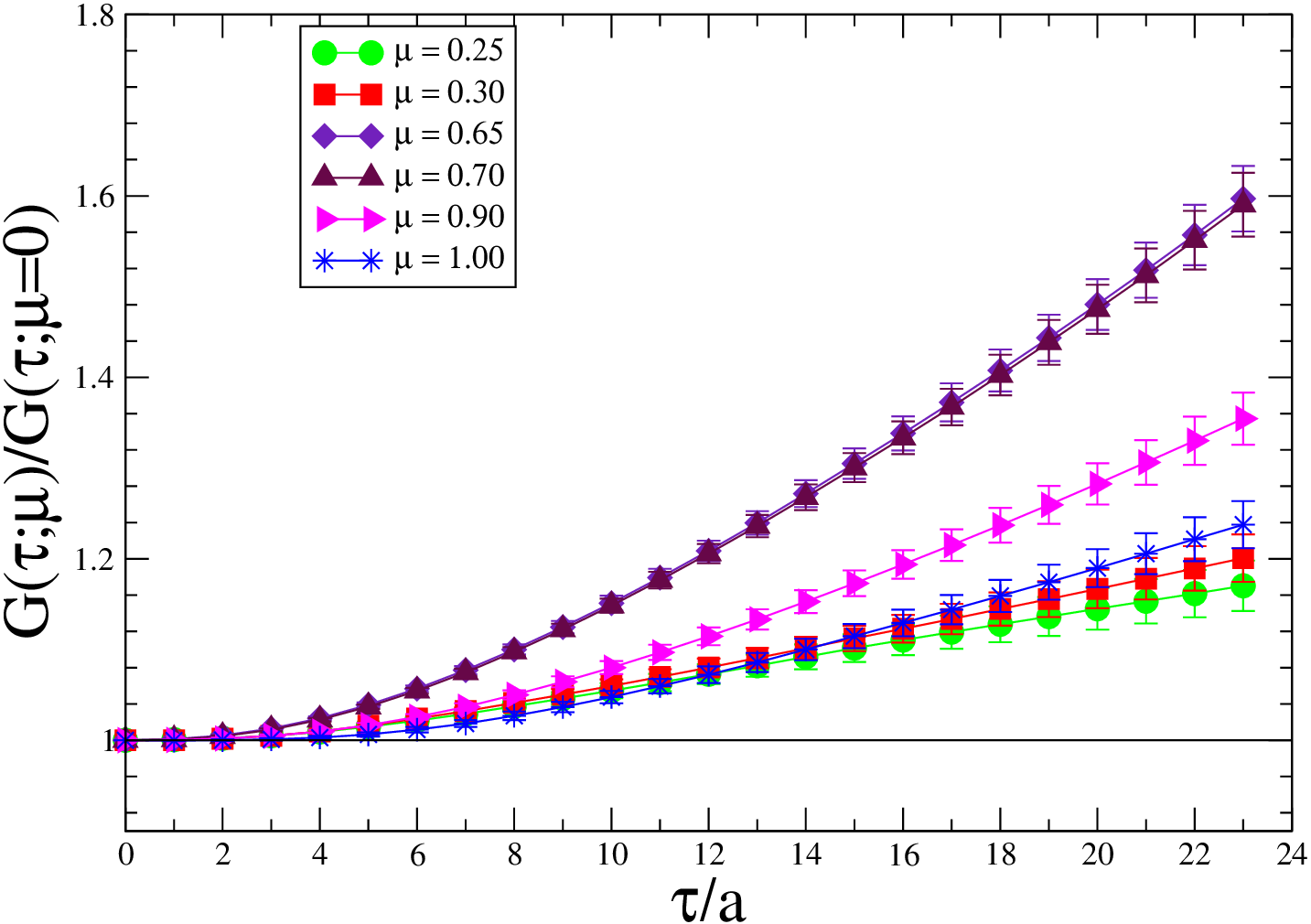}
\end{center}
\caption{(left) The ratio $\sum_\xv G(\xv,\tau;\mu)/\sum_\xv G(\xv,\tau;0)$
  for $^1P_0$ correlators on $12^3 \times 24$ with $Ma = 5.0$. Due to the noisiness of
  the P-wave data, only a limited $\tau$ range is shown; (right) the
  corresponding ratio for $^1S_0$ correlators for comparison.}
\label{fig:pwave}
\end{figure}
In contrast to the $S$-wave states, it is difficult to find stable
exponential fits to the $P$-wave correlators with the current
Monte-Carlo data before statistical noise sets in, except for the case
$\mu a \le 0.25$. We therefore use a more primitive quantity: the ratios of the
$^1P_0$ state correlators at several values $\mu\not=0$ (chosen within the three
different regimes) to the correlator at $\mu=0$ are compared in
Fig.~\ref{fig:pwave}. The other P-wave states (e.g., $^3P_1$ state) show similar behaviors. The corresponding ratios of the $^1S_0$ state
correlators are shown to the right.  Note that any effect we observe
is entirely due to the dense medium.

The $S$-wave correlators
may be represented as a sum of exponential functions $\sum_i
A_ie^{-\Delta E_i\tau}$ with
$\Delta E_1 < \Delta E_2 < \Delta E_3 \cdots$ so that the large-$\tau$ behavior is dominated
by the lowest energy state $\Delta E_1$. The $S$-wave correlator ratios in
Fig.~\ref{fig:pwave} confirm this expectation. At large $\tau$ the
ratio will be simply $R(\tau;\mu)=e^{-\delta(\mu)\tau}$ where
  $\delta(\mu)=\Delta E_1(\mu)-\Delta E_1(\mu=0)$.  In the BEC region
($\mu \le \mu_1$) and the deconfined region ($\mu \ge \mu_2$), the
ratio is $\sim 20$\% at $\tau/a = 23$ and may be
approximated as a straight line,
$R(\tau;\mu)\approx1-\delta(\mu)\tau$, consistent with the small,
negative $^1S_0$ state energy difference $\delta(\mu)$ that was previously
observed. For the BCS region ($\mu_1 \ge \mu a \ge \mu_2$), the ratio is
$\sim 60$\% at $\tau/a = 23$ and may be approximated as an
exponential function, which is consistent with the large 
energy difference in this region. In either case the ratio increases
monotonically with $\tau$.

Unlike the simple behavior seen for the $S$-wave,
the $P$-wave correlator ratios show an interesting
$\tau$-dependence. In the BEC and BCS regions ($\mu \le
\mu_2$), the $P$-wave correlator ratios behave similar to the $S$-wave,
but in the deconfined region ($\mu \ge \mu_2$),
the $P$-wave correlator ratios are non-monotonic, initially decreasing with
$\tau$ before turning to rise above unity for $\tau a\sim4$. On the
other hand, the ratios of $P$-wave state correlators on $12^3 \times 16$ and $16^3 \times 12$ lattices show monotonic behavior as the $S$-wave correlator
ratios do, which suggests a subtle interplay of density and 
temperature effects on the $P$-wave states.

\section{Discussion}
\label{sec:rampant_speculation}

We have mapped out the variation of heavy quarkonium $S$-wave
states in a two-color baryon-rich medium with $\mu/T\gg1$, as both $\mu$ and $T$ are
varied. The behaviour is unexpectedly complex; the state energies initially
decrease, then plateau, and finally rise again to become comparable or even
exceed the vacuum value. The medium effect increases as the heavy quark mass $M$
is decreased. Using the string tension to set the scale we find a downwards energy shift 
$\Delta E\sim40$MeV for $Ma=3.0$ and $\mu_1<\mu<\mu_2$.

It is natural to seek an explanation in terms of the various ground states set
out in \cite{Hands:2006ve, Hands:2010gd}. Initially assume $T\approx0$;
$\Delta E(\mu)$ must then arise solely from interactions between the heavy quark
pair $QQ$ (or equivalently $Q\bar Q$) and light quarks $q$ in the medium. An
obvious possibility is the formation of two $Qq$ states, or conceivably even a
tetraquark $QqqQ$. In vacuum the quarkonium state usually lies ${\cal
O}(\Lambda_{\rm QCD})$ below the threshold for this to
occur~\cite{Brambilla:2002nu}. In a baryonic medium (ie.  above onset
$\mu>\mu_o$) the $q$s are already present and no longer need to be excited from
the vacuum, so that the $QQ$ energy may now be above threshold. 
A naive energy budget must take into account both the breaking apart
of the gauge singlet $QQ$ and $qq$ states, and the subsequent formation of two
$Qq$ states. We assume that only $qq$ breaking has any significant
$\mu$-dependence, since it is the properties of the $q$-medium that evolve with
$\mu$.

Henceforth assume that $\mu_1$ coincides with the transition from BEC to
quarkyonic phase at $\mu_Q$ identified in \cite{Hands:2010gd}, and $\mu_2$ with
the deconfining transition at $\mu_D$. For $\mu_o<\mu<\mu_1$ the medium thus
consists of tightly-bound diquark states, which are also Goldstone bosons
associated with superfluidity, with mass 
proportional to $\surd j$~\cite{Kogut:2000ek,Hands:2007uc}. 
We deduce that the energy required to break
such bound pairs falls as the quark density $n_q$ rises and, since the effect
increases as $M$ falls, with the ultimate
separation of the resulting $q$ -- $q$ system. Both these factors suggest an in-medium 
screening of the interaction between light quarks, or equivalently a non-trivial 
$\mu$-dependence of the Goldstone decay constant $F_\pi(\mu)$, whose detailed mechanism
remains unclear. 

For $\mu_1<\mu<\mu_2$ the system is hypothesised to be in a quarkyonic
phase, which we take to be a state in which quarks, though still confined, form
a degenerate system with well-defined Fermi energy $E_F\sim\mu$, and in which
superfluidity arises through BCS condensation of weakly-bound and spatially
delocalised Cooper pairs. To excite light $q$s capable of
forming $Qq$ states now thus requires an energy of ${\cal
O}(\Delta)$, the superfluid gap, believed to be approximately $\mu$-independent
in this regime~\cite{Hands:2010gd}. Forming a $Qq$ state at rest also requires
the heavy quark to have kinetic energy ${\cal O}(\mu^2/M)$, which is a small,
perhaps negligible, correction. This accounts for the approximate
$\mu$-independence of $\Delta E$ observed in this regime. Since the $q$s are no
longer bound within Goldstone bosons, we expect their excitation energy in this
regime to vary
linearly with $j$, which is thus responsible for the mild increase of $\Delta E$
with $j$ seen in Fig.~\ref{fig:qq}.

Finally, for $\mu>\mu_2$ the system is deconfined, as signalled by the
non-vanishing expectation of the Polyakov loop. In this regime 
the physical states could in principle be isolated heavy quarks dressed by a cloud of both
light quarks and now gluons, although we should not at this stage rule out the
persistence of bound states -- indeed, the contrast between $S$- and $P$-wave
states, which are more spatially extended, in this region in
Fig.~\ref{fig:pwave} suggests
an interesting story remains to be told. In either case
the light constituents are now expected to
have a Debye mass $m_D\propto g\mu$ generated via quantum loop corrections; hence
$\Delta E$ now rises with increasing $\mu$, as confirmed by Fig.~\ref{fig:1S0muT}.
Thermal mass generation $m_D\propto gT$ may also be responsible for the
systematic increase of $\Delta E$ with $T$ seen in Fig.~\ref{fig:1S0muT}; it is
notable in this case that the thermal effect appears to be equally manifest in all
three $\mu$-regimes.

The reader will no doubt agree that these are speculative ideas,
inevitably constrained by theoretical pictures which can only become
accurate in limits of vanishingly small or asymptotically high densities. It may
well turn out that our identification of three different $\mu$-regimes is
over-elaborate and masks a more unified explanation. Nonetheless,
in view of the interesting heavy quarkonia physics in the non-zero temperature
environment\cite{Aarts:2010ek,Aarts:2011sm}, we expect from this exploratory study
that heavy quarkonia can yield important insights into the nature of
dense baryonic matter.

\section*{Acknowledgements}

This work is carried out as part of
the UKQCD collaboration and the DiRAC Facility jointly funded by STFC, the 
Large Facilities Capital Fund of BIS and Swansea University.
We thank the DEISA Consortium (www.deisa.eu), funded 
through the EU FP7 project RI-222919, for support
within the DEISA Extreme Computing Initiative.
SK is grateful to STFC for a Visiting
Researcher Grant and is supported by the National Research Foundation
of Korea grant funded by the Korea government (MEST)
No.\ 2011-0026688. JIS is supported by Science Foundation Ireland
grant 11-RFP.1-PHY3193.

\end{document}